\documentclass[12pt,preprint]{aastex}

\newcommand{\ps}{\mbox{$P_{\rm s}$}}
\newcommand{\pb}{\mbox{$P_{\rm orb}$}}
\newcommand{\rn}{\mbox{$R_{\rm m}$}}
\newcommand{\rc}{\mbox{$R_{\rm c}$}}
\newcommand{\my}{\mbox{$M_{\odot}{\rm yr}^{-1}$}}
\newcommand\lsim{\lower0.5ex\hbox{$\; \buildrel < \over \sim \;$}}
\newcommand\gsim{\lower0.5ex\hbox{$\; \buildrel > \over \sim \;$}}

\title
{EXPLORATION OF THE $\ps-\pb$ RELATION FOR WIND-FED X-RAY PULSARS
}

\author
{Hai-Lang Dai, Xi-Wei Liu and Xiang-Dong Li}

\affil{Department of Astronomy, Nanjing University,
    Nanjing 210093, P. R. China}

\email{hldai@nju.edu.cn; lixd@nju.edu.cn; liuxw@nju.edu.cn}

\date{}

\begin{document}

%\maketitle

\begin{abstract}
We have investigated the relation between the orbital periods
($\pb$) and the spin periods ($\ps$) of wind-fed X-ray pulsars in
high-mass X-ray binaries (HMXBs), based on population synthesis
calculations of the spin evolution of neutron stars during the
pre-HMXB stage. We show that most of the neutron stars either have
steady accretion or still reside in the radio pulsar phase when
the donor star starts evolving off the main sequence. In the
former case the values of $\ps$ can be decelerated to be $\sim
10^2-10^3$ s depending on $\pb$.
%Our results are in rough agreement with the distributions of HMXBs
%in the $\ps-\pb$ diagram.
We briefly discuss the possible origin of the $\ps-\pb$
correlation in Be/X-ray binaries, and the existence of HMXBs with
main sequence donors. We have also investigated the evolution of
the inclination angle between the magnetic and spin axes of
neutron stars in a massive binary, suggesting secular alignment of
the magnetic and spin axes during their evolution.
\end{abstract}


\section{INTRODUCTION}
High-mass X-ray binaries (HMXBs) can be roughly divided into two
types: a supergiant star and a Be star are contained respectively.
The X-ray source often is a pulsar and powered by accretion of the
material offered by the companion star. In the supergiant systems,
either Roche-lobe or stellar wind accretion occurs, while in the
Be systems commonly only the latter process takes place since the
Be star is well inside its Roche lobe \citep{tau05}.

The relation between the spin periods $\ps$ and the orbital
periods $\pb$ of HMXBs was first studied by
\citet{cor84,cor85,cor86}, who pointed out that there may be a
$\ps\propto P_{\rm orb}^{4/7}$ correlation for neutron stars in
systems with radially expanding winds, while $\ps\propto P_{\rm
orb}^{2}$ for Be systems. \citet{van87} noticed that the spin and
orbital periods agree reasonably well with the former correlation
for the supergiant systems in which the X-ray source is powered by
stellar wind accretion. The $\ps-\pb$ relation has been studied by
\citet{ste86} in terms of the equilibrium spin period, where the
corotation radius equals to the magnetospheric radius defined by
equating the ram pressure of accreting matter with the magnetic
pressure of the dipole field of the neutron star
\citep[e.g.,][]{dav73,lam73}. However, in order to obtain
quantitative agreement between the predicted and observed spin
periods for given orbital periods, the assumption of the
equilibrium spin periods requires a mass accretion rate two orders
of magnitude lower than observed for supergiant systems
\citep{ste86}. \citet{wat89} presented a comprehensive explanation
for the $\ps-\pb$ relation for HMXBs. For supergiant systems they
suggested that the difference between the required and the
observed mass accretion rates can be explained by assuming that
the present-day periods are the equilibrium spin periods for the
stellar wind of the optical star when it was still on main
sequence (MS). The $\ps-\pb$ correlation for Be/X-ray binaries was
suggested to result from the equilibrium spin realized by
interaction between the neutron star magnetic field and the Be
star's equatorial wind - the wider orbits expose to the neutron
stars to a lower wind density on average and hence lower accretion
rate. The latter point was criticized by \citet{king91}, who
proposed that the observed $\ps-\pb$ relation was probably the
result of an earlier evolutionary stage, as for supergiant
binaries. By calculating the angular momentum carried by the Be
star's wind material, \citet{li96} showed that for Be/X-ray
binaries in relatively narrow orbits ($\pb\lsim$ 100 days), the
equilibrium spin could be reached by angular momentum transfer via
a disk, formed in the equatorial wind of the Be stars; in wider
binaries, the low angular momentum of the wind matter prevents the
existence of an accretion disk, and the neutron star's spin
remains close to its previous equilibrium value at an earlier
evolutionary stage. \citet{zha04} have calculated the detailed
spin evolution of a wind-fed neutron star in OB binaries prior to
the HMXB phase, by simulating the time-development of the mass
loss rate and radius expansion of a 20 $M_{\odot}$ donor star.

The above-mentioned investigations are usually either
semi-analytic or numerical only for individual cases. To better
understand the distribution of HMXBs in the $\ps-\pb$ diagram, an
evolutionary population synthesis incorporated with neutron star
spin evolution is needed. In the present paper, we have conducted
a Monte Carlo population synthesis study of the spin evolution of
a neutron star in a massive binary. Due to the large theoretical
uncertainties on the formation and evolution of HMXBs, our work is
limited to the spin evolution of neutron stars before the HMXB
phase. The theoretical considerations are described in \S 2. The
calculated results are presented in \S 3 and their implications
for accretion processes in HMXBs are discussed in \S 4.

\section{MODEL}
\subsection{Spin evolution}

We consider a $1.4 M_{\odot}$ magnetized neutron star in binaries
with a massive MS companion star. We adopt a simplified version of
the theoretical model outlined by \citet{dav81} to calculate the
neutron star's spin evolution before steady accretion occurs, as
briefly described below.

\noindent{\em Case a: the pulsar phase}

The newborn neutron star first appears as a rapidly rotating,
radio pulsar after its birth in a supernova explosion, provided
that its radiation is strong enough to expel the wind material
coming from the companion star outside the radius of the light
cylinder $r_{\rm lc}=c\ps/2\pi$ or the Bondi accretion radius
$r_{\rm G}=2GM/v_{\infty}^2$ \citep{bon44}, where $G$ is the
gravitational constant, $M$ the mass of the neutron star and
$v_{\infty}=10^8v_8$\,cms$^{-1}$ the relative wind velocity at the
neutron star's orbit. The neutron star spins down due to magnetic
dipole radiation and/or particle emission,
\begin{equation}
I\dot{\Omega}_{\rm s}=-\frac{2}{3}\frac{\mu^2\Omega_s^3}{c^3},
\end{equation}
where $I$ is the moment of inertia, $\mu=10^{30}\mu_{30}$\,Gcm$^3$
the magnetic dipole moment, and $\Omega_{\rm s}$ the angular
velocity of the neutron star, respectively.

The pulsar phase will break down in two distinct circumstances.

(i) The pulsar phase will cease when the wind material penetrates
inside $r_{\rm lc}$, and the resulting $\ps$ is derived by
balancing radiation pressure from the pulsar with the stellar wind
ram pressure at $r_{\rm lc}$ \citep{dav81},
\begin{equation}
P_{\rm a} \simeq 0.8 \mu_{30}^{1/3} \dot{M}_{15}^{-1/6}
(M/M_{\odot})^{1/3}v_8^{-5/6}\,{\rm s},
\end{equation}
where $\dot{M}=10^{15}\dot{M}_{15}$ \,gs$^{-1}$ is the mass flow
rate onto the neutron star.

(ii) The pulsar phase will end when the pressure gradient become
important as the outer boundary $R_{\rm a}$ of the envelope where
the density $\rho\simeq \rho_{\rm \infty}$ approaching to $r_{\rm
G}$, i.e., $R_{\rm a}\simeq r_{\rm G}$, and the corresponding
$\ps$ is \citep{dav81},
\begin{equation}
P_{\rm a}\simeq
1.2\dot{M}_{15}^{-1/4}\mu_{30}^{1/2}v_8^{-1/2}\,{\rm s}.
\end{equation}

%The pulsar phase will cease when the wind material penetrates
%inside either $r_{\rm lc}$ or $r_{\rm G}$. The resulting $\ps$ is
%then \citet{dav81},
%\begin{equation}
%P_{\rm a} = 0.8 \mu_{30}^{1/3} \dot{M}_{15}^{-1/6}
%(M/M_{\odot})^{1/3}v_8^{-5/6}\,{\rm s},
%\end{equation}
%or
%\begin{equation}
%P_{\rm a}=1.2\dot{M}_{15}^{-1/4}\mu_{30}^{1/2}v_8^{-1/2}\,{\rm s},
%\end{equation}
%where $\dot{M}=10^{15}\dot{M}_{15}$ \,gs$^{-1}$ is the mass
%accretion rate.

\noindent{\em Case b: the propeller phase}

The propeller phase follows the pulsar phase. During this period
the magnetospheric radius $\rn=[\mu^4/(2GM\dot{M}^2)]^{1/7}$, is
larger than the corotation radius $\rc=(GM/\Omega_{\rm
s}^2)^{1/3}$. The plasma interacts with the neutron star
magnetosphere, but further accretion is inhibited by the
centrifugal barrier, and the infalling matter is accelerated
outward, taking away the angular momentum of the neutron star.
Though the propeller effect has be investigated extensively, there
exist large uncertainties in the efficiency of angular momentum
loss during the propeller regime
\citep[e.g.][]{pr72,is75,dav81,wan85,ik01}. Here we assume that
the infalling material is ejected with the corotating velocity at
$\rn$, and the spin-down torque is \citep{wan85,jia05}
\begin{equation}%\label{}
N=I\dot{\Omega}_{\rm s}=-\dot{M}R_{\rm m}^2\Omega_{\rm s}.
\end{equation}
The typical spin-down time-scale $\tau=|\Omega_{\rm
s}/\dot{\Omega}_{\rm s}|$ can be estimated to be
\begin{equation}
\tau\simeq 2.2\times
10^4\mu_{30}^{-8/7}\dot{M}_{15}^{-3/7}(M/M_{\odot})^{2/7}I_{45}\,{\rm
yr}.
\end{equation}
The spin-down process ends until $\ps$ reaches the equilibrium
spin period
\begin{equation}
P_{\rm eq}\simeq
17\mu_{30}^{6/7}\dot{M}_{15}^{-3/7}(M/M_{\odot})^{-5/7}\,{\rm s},
\end{equation}
and we assume that steady accretion takes place afterwards
\citep[see however,][]{al76,el76}.

The spin period of the neutron star may be further changed after
$P>P_{\rm eq}$ in the following accretion phase. But we stop the
calculations when either $P_{\rm eq}$ is reached within the MS
lifetime or the companion star evolves off the MS (so that strong
wind accretion or Roche lobe overflow occurs) \citep[see
also][]{wat89}. In this work we do not consider narrow HMXBs with
Roche-lobe overflow (like SMC X$-$1, Cen X$-$3, and LMC X$-$4),
where the neutron stars are most likely to accrete from an
accretion disk. For wind-fed systems like Vela X$-$1, numerical
calculations \citep[e.g.][]{fry88,mat92,anz95,ruf99} have
suggested that there are no significant angular momentum transfer
onto the neutron star when radially-expanding wind material
accretes onto the neutron star. This may result in only small
deviation from the instantaneous (equilibrium if reached) spin
periods at the beginning of the accretion phase. {\em CGRO/BATSE}
observations have shown a random walk in their spin frequencies
with alternating spin-up and spin-down \citep{bil97}. The spin
evolution in Be/X-ray pulsars has a completely different picture,
and will be discussed separately.

\subsection{Evolution of the inclination angle}
Besides the spin evolution, we have considered the evolution of
the inclination angle $\chi$ between the spin and magnetic axes in
wind-fed neutron stars. This part of work is partially motivated
by \citet{bul03}. By analyzing the light curves of 89
accretion-powered pulsars these authors obtained an upper limit on
the inclination angles $\chi<50\degr$, suggesting that the
magnetic axis tended to be aligned with the rotation axis. Similar
analyzes and conclusions have also been conducted by
\citet{wan81}, \citet{lea91}, and \citet{lea95}. More recently
\citet{bla05} presented the contemporaneous high-energy and radio
observations of the HMXB 4U 2206+54 conducted with
\textit{INTEGRAL} and VLA, which firmly indicate that 4U 2206+54
hosts a magnetic accreting neutron star. The absence of pulsations
are most likely to be due to geometrical effect that the
inclination angle of the neutron star is very small so that no
pulsation can be detected.

\citet{wan82} show that the inclination angle in a binary X-ray
pulsar will vary as a result of its interaction with the
surrounding plasma, becoming smaller (larger) during spin-down
(spin-up). According to Eqs. (27), (36c), and (37c) in \citet
{wan82}, we can derive the changing rate of $\chi$ as follows,
\begin{eqnarray}
\dot{\chi}&=&\left\{
\begin{array} { ll}
  N\cot\chi/(I\Omega_{\rm s}),&\ \chi\gtrsim\theta_{\rm c}\\
  3N\sin\chi/(I\Omega_{\rm s}\sin^2\theta_{\rm c}),&\
  \chi\lesssim\theta_{\rm c}
\end{array}
\right.
\end{eqnarray}
where $\theta_c$ is the critical value of the meridional angle
measured from the north magnetic pole.

\subsection{Evolution of the mass flow rate onto neutron star}

We employed an evolutionary population synthesis method to
investigate the distribution of the orbital periods and companion
masses for the natal neutron star binaries (i.e. at the moment
when the neutron star was born). We started with a large set of
primordial binaries and generated the systems that evolved to
contain a neutron star and a MS companion. The initial mass
function of \citet{Kro93} was taken for the masses of the primary
star (the progenitor of the neutron star of mass $M_1$). For the
secondary star (of mass $M_2$), we assumed a uniform distribution
of the mass ratio for $0<q\equiv M_2/M_1\le 1$. A uniform
distribution of $\ln a$ was also taken for the binary separation
$a$.  We adopted the binary population synthesis code developed by
\citet{hur00,hur02} to evolve the primordial binaries. This code
incorporates evolution of single stars with binary-star
interactions, such as mass transfer, mass accretion,
common-envelope (CE) evolution, collisions, supernova kicks, tidal
friction and angular momentum loss mechanisms.  Most of our
adopted parameters are the same as the standard ones described in
\citet{hur02}.  The star formation rate parameter is $S = 7.6085$
yr$^{-1}$, corresponding to a rate of $\sim 0.02$ yr$^{-1}$ for
core-collapse supernovae (SNe) in our Galaxy, assume all the stars
with masses $> 8M_{\odot}$ die through SNe. During the SN
explosions, we apply a Maxwellian distribution in the kick
velocities with a mean of 265 kms$^{-1}$ imparted on the newborn
neutron stars \citep{hobb05}. The treatment of Roche-lobe overflow
(RLOF) mass transfer in the primordial binary is presented
specifically in \citet{hur02}, and here we describe the stability
criterion of mass transfer briefly. Mass transfer via Roche-lobe
overflow takes place on either nuclear, thermal, or dynamical
time-scales, depending on whether the primary remains in thermal
equilibrium as it loses mass, and the radius of the primary
increases faster than the Roche-lobe. Stars with deep surface
convective zones, e.g. giants or naked helium giants, are
generally unstable to dynamical timescale mass loss and will enter
a CE evolution. The stable mass accretion rate of the secondary
star is limited by Eddington accretion rate. Actually, the
secondary may be spun up and become a Be star when it accretes
enough mass. This circumstance however, is not included in our
consideration since the origin of Be phenomena is still unclear
and it is difficult to model the mass transfer processes in
Be/X-ray binaries. The CE parameter $\alpha$ was set to be 1 as a
typical value and we varied it from 0.1 to 2 in the calculations
\citep{dew00, tau01}. Our product is a set of ``incipient" neutron
star in massive binaries with a particular distribution of $\pb$
and $M_{2}$ (shown in Fig.~1). Other binary parameters, such as
the radii, surface temperatures and luminosities of the companion
stars can also be obtained. These parameters were then used to
evaluate the mass loss rates from the companion stars and the mass
flow rates onto the neutron stars.

For comparison with the observed properties of HMXBs, our
calculations are limited to systems with $\pb<1000$ days and
$10M_{\odot}\lsim M_2\lsim 30M_{\odot}$. The mass loss rate
$\dot{M}_2$ was estimated with the prescription described by
\citet{nie90},
\begin{equation}
-\dot{M}_2=9.6\times10^{-15}
R_2^{0.81}L_2^{1.24}M_2^{0.16}\,M_{\odot}{\rm yr}^{-1},
\end{equation}
where $R_2$ and $L_2$ are the radius and luminosity of the
companion star. All the quantities in Eq.~(8) are evaluated in
solar units. Assume that the stellar wind expands isotropically at
a speed of $v_{\rm w}$, the wind density $\rho_{\rm w}$ at the
orbit of the neutron star is
\begin{equation}
\rho_{\rm w}=-\dot{M}_2/(4\pi a^2v_{\rm w}),
\end{equation}
and the mass flow rate onto neutron star is roughly given by
\citep{bon44}
\begin{equation}
\dot{M}=\pi r_{\rm G}^2\rho_{\rm w}v_{\infty}.
\end{equation}

\section{RESULTS}

We have done calculations of the spin and inclination evolutions
for $6\times10^5$ neutron star binary systems based on the
theoretical model presented in \S 2. For the initial neutron star
magnetic fields $B$ we assumed that $\log B$ is distributed
normally with a mean of 12.5 and a standard deviation of 0.3. No
field decay is considered. The initial distribution of the
inclination angle $\chi$ is randomly distributed in the range of
$(0,\pi/2)$. We set the initial spin periods to be distributed
uniformly between $10$ and $100$ ms, though they have little
influence on the final results. We stop our calculations when
either $\ps$ reaches $P_{\rm eq}$ or the companion star begins to
evolve off the MS with the arguments afore.

The calculated results are presented in Figs.~2$-$5. Panels (a)
and (b) in Figs.~2$-$3 correspond to the relative wind velocity at
the neutron star's orbit $v_8=1$ and 2 respectively. Figure 2
shows the final distribution of all the binaries in the $\ps-\pb$
diagram. The relative numbers of binary systems are displayed
sketchily with the darkness of the shaded regions. According to
our calculation, we find that, when $v_8=1$, around $68\%$ neutron
stars can reach the equilibrium periods to allow wind accretion
within the MS lifetime $t_{\rm MS}$ , while others would still be
in either the pulsar ($\sim 31\%$) or the propeller ($\sim 1\%$)
phases. This can be clearly seen in Fig~2, where the former two
groups of neutron stars occupy the distinct upper and lower shaded
regions. The small number of stars in the propeller phase
originates from its much shorter duration \citep{dav81}. If $v_8$
is increased to be 2, the mass flow rates onto the neutron stars
will be lower by a factor of $\sim 16$ according to Eqs.~(9) and
(10), further extending the spin-down time in the pulsar phase. So
the corresponding numbers become $\sim 42\%$, 55\% and 3\%,
respectively. The larger wind velocity also induces longer
equilibrium periods. Figure 2 reveals that up to half of neutron
stars in binary systems cannot reach the equilibrium periods,
indicating that there might be hundreds of ``sleeping" neutron
star $+$ MS star binaries like PSR B1259$+$63 in the Galaxy (see
also Fig.~4 below). They could be observed as radio pulsars or in
X-rays depending on whether the interaction between the neutron
star and the companion's winds becoming active. Most of them have
$\ps\sim 1$ s and $\pb$ from tens of to hundreds of days.

To compare the calculated results with observations of HMXBs, we
show the distribution of those neutron star binaries with $\ps\geq
P_{\rm eq}$ and of the observed HMXBs in Fig.~3. Asterisks and
diamonds mark the Be and supergiant wind-fed HMXBs respectively,
and crosses for Roche-lobe overflow systems \citep[data are taken
from][]{rag05}. We first discuss the persistent, supergiant
systems. Obviously the spin periods of neutron stars can approach
$P_{\rm eq}$ if the total spin-down timescale $t_{\rm spin}$ is
less than the MS lifetime $t_{\rm MS}$ of the companion star. As
we argued before, for this type of systems, the current spin
periods are likely to be around the equilibrium periods $P_{\rm
eq}$ (MS) attained during the MS stage because of the inefficient
angular momentum transfer in the accretion phase. For these
systems to be observed as HMXBs, the accretion rate must have been
increased to be high enough that the spin periods are larger than
the (current) equilibrium period $P_{\rm eq}$ (sg) when the
companion star becomes a supergiant \citep{ste86,wat89}, which is
plotted with the dotted line in Fig.~3 according to Eq.~(6). Here
we take $M_2=20M_{\odot}$, $\dot{M}_2=10^{-6}\,\my$, and
$\mu_{30}=3$ \citep{cob04} as typical values of the supergiant
systems. This period roughly serves as the lower limit for $\ps$
of the neutron stars in supergiant HMXBs. Moreover, since the mass
accretion rates generally decrease with $\pb$, only those in
narrow orbits ($\pb\lesssim 15$ d) can have X-ray luminosities
$\gtrsim 10^{36}$ ergs$^{-1}$ as observed.

%The reason that we consider both $P_{\rm eq}$ and $P_{\rm br}$ is
%that the latter was derived under the assumption that there is a
%static quasi-spherical envelope around the neutron star
%\citep{dav81}, which may not be appropriate for accreting X-ray
%pulsars in supergiant HMXBs, which are usually surrounded by a
%small disk even in wind-accretion phase \citep{fry88}. Similar
%situations exist in Be/X-ray binaries. Actually we can see from
%Fig.~4 that $P_{\rm eq}$ in accretion phase may be more suitable
%as the lower limit of $\ps$ for both types of systems.

Be/X-ray binaries generally have much longer orbital periods than
the supergiant systems, and are usually transient X-ray sources
observable during outbursts. The structure of the Be star's winds
is quite complicated, consisting of a relatively dense, slowly
expanding, disk-like, equatorial wind and a fast, isotropically
expanding, polar wind \citep{wat89}. \citet{wat89} suggested that
the $\ps-\pb$ correlation for Be/X-ray binaries could be explained
by setting $\ps=P_{\rm eq}$ obtained in the disk winds. However,
current understanding of the Be disk winds \citep{oka01} is very
different from the radial wind model used by \citet{wat89}. The
evolution of the Be star's disk and the high eccentricities of the
orbits also mean that the mass flow rate onto neutron stars in Be
systems are always highly variable on both long-term and
short-term timescales. Because of these complications we did not
calculated the spin evolution of neutron stars accreting from the
disk winds, but present constraints on their possible locations in
the $\ps-\pb$ diagram instead. We first note that the calculated
$P_{\rm eq}$ (MS), under the assumption of isotropic winds, can be
regarded as the upper limits for $\ps$ of Be/X-ray binaries, since
$\dot{M}$ is much lower from the polar winds than from the disk
winds (these values should be taken as the averaged ones since we
have ignored the eccentricity of the Be star binaries). Then,
similar as in the supergiant systems, we can set the lower limit
of $\ps$ for Be/X-ray binaries to be the equilibrium period
$P_{\rm eq}$ (disk) in outbursts when the the disk winds dominate
accretion. From the observed data of 36 Be/X-ray binaries compiled
by \citet{rag05}, we have derived a correlation between the
maximum luminosities and the orbital periods,
\begin{equation}
\log (L_{\rm X,max}/10^{35}\,{\rm ergs}^{-1})=4.53(\pm 0.66)
-1.50(\pm 0.33)\log(\pb/{\rm d}).
\end{equation}
Insert Eq.~(11) and $\mu_{30}=3$ into Eq.~(6) we obtain
\begin{equation}
\log P_{\rm eq}({\rm disk})\simeq -0.29+0.64\log(\pb/{\rm d}),
\end{equation}
%\begin{equation}
%\log P_{\rm br}({\rm e})=-0.29+1.1\log(\pb/{\rm d}).
%\end{equation}
which is plotted in the solid line in Fig.~3. We see that a large
fraction of Be/X-ray binaries lie between $P_{\rm eq}$ (MS) and
$P_{\rm eq}$ (d). The peculiar location of A 0535$-$669 may be due
to its relatively low magnetic field.

The cumulative luminosity distributions of HMXBs based on our
population synthesis calculations are shown in Fig.~4 with
different values of the parameters $\alpha$, $v_8$, and $\sigma$.
The results show that change of these parameters do not
significantly influence the final outcome, which are roughly
compatible with the observed luminosity distributions of HMXBs in
our Galaxy \citep{gri02}.

In Fig.~5 we plot the inclination angle distribution for X-ray
pulsars in HMXBs. It is easily seen that most of the inclination
angles are within 1 rad, in general agreement with the analysis of
the observation data \citep{bul03}. The evolution of the
inclination angle seems to be insensitive to the values of the
relative wind velocity.

\section{DISCUSSION}
We have calculated the spin evolution of neutron stars in a
massive binary. The main ideas in this work are of two folds.
First, to appear as X-ray binaries, the spin periods of neutron
stars should be longer than the current (instantaneous)
equilibrium periods. Second, to satisfy this condition, the
neutron stars should be spun-down enough during the previous
propeller phase when the companion stars had relatively weak
winds. These two requirements set constraints on the spin period
distribution in wind-fed X-ray pulsars. Since $\dot{M}$ depends on
the orbital separations, this may result in the particular
distributions of the supergiant and Be HMXBs in the $\ps-\pb$
diagram. The numerical results presented in Figs.~2 and 3, roughly
accounting for the main features of the $\ps-\pb$ distribution of
HMXBs, lend support to the original suggestions by \citet{cor84},
\citet{ste86}, and \citet{wat89}. Our calculations of the
alignment during spin-down imply that most HMXBs may have
relatively small inclination angles.

We note that our preliminary results are subject of lots of
uncertainties and simplified treatment adopted. In case of
isotropic wind accretion, a considerable fraction of neutron stars
can reach $P_{\rm eq}$ when the companion star is still on the MS.
The detail number is determined not only by the binary
evolutionary processes, but also critically by the propeller
mechanism. The spin-down torque (Eq.~[4]) adopted in this work is
among the most efficient ones \citep[][and references
therein]{jia05}, and the results should be taken as to be the most
optimistic. For less efficient spin-down torques ones, the number
of binaries in the propeller phase will obviously increase due to
the longer spin-down timescales. The magnitude of $P_{\rm eq}$
depends on the mass loss rate, the wind velocity, and the magnetic
field strength of the neutron star. With the mass loss rates given
by Eq.~(9) and typical magnetic fields of $\sim 3\times 10^{12}$
G, the neutron star's spin can be decelerated to be $\sim
10^2-10^3$ s within $t_{\rm MS}$ of the secondary star. The very
long period $\ps=10^4$ s of 2S 0114$+$650 \citep{hal00}, however,
may be explained by a ultra-strong initial magnetic field ($B\gsim
10^{14}$ G), in order to be spun-down efficiently by the propeller
effect \citep{li99}.

The structure of the Be star's winds is much more complex and
variable than in supergiant systems, and the simple propeller
spin-down may not be applicable to this type of systems. It is
likely that the $\ps-\pb$ distribution and correlation of Be/X-ray
binaries result from the balance between the spin-up during
outbursts and spin-down during quiescence. Assume that the spin-up
and spin-down torques are $\dot{M}_{\rm o}(GMR_{\rm mo})^{1/2}$
and $-\dot{M}_{\rm q}R_{\rm mq}^2\Omega_{\rm s}$ respectively, we
can derive the following expression for the equilibrium period
\citep[see also][]{men99},
\begin{equation}
P_{\rm eq}=2\pi(\frac{GM}{R_{\rm mo}^3})^{-1/2}(\frac{\dot{M}_{\rm
q}R_{\rm mq}^2}{\dot{M}_{\rm o}R_{\rm mo}^2})(\frac{t_{\rm
q}}{t_{\rm o}})\propto \dot{M}_{\rm o}^{-3/7}(\frac{\dot{M}_{\rm
q}}{\dot{M}_{\rm o}})^{3/7}(\frac{t_{\rm q}}{t_{\rm o}}).
\end{equation}
Here the subscripts o and q denote quantities evaluated during
outbursts and quiescence, respectively. All the three terms on the
right hand side of Eq.~(13) are likely to increase with $\pb$,
which might account for the observed $\ps-\pb$ correlation of
Be/X-ray binaries.

Our calculated results also indicate the existence of HMXBs with MS
donors in the Galaxy. These systems, as the progenitors of
supergiant HMXBs, are less luminous than the latter (but with
similar spin periods). Figure 4 suggests that there could be a few
hundred of these sources in the Galaxy with X-ray luminosities
ranging from $\sim 10^{33}$ to $10^{35}$ ergs$^{-1}$ (If the
propeller effect is not considered, the total number of X-ray
binaries, most of which are Be/X-ray binaries, can reach a few
thousand).
%however, the numbers of HMXBs with MS donors not only in high
%luminosity but in low luminosity are much larger than the numbers
%of HMXBs with supergiant donors according to the results of
%population synthesis. The results in Fig. 4. are consistent well
%with the observations \citep{gri02}, which indicate that there are
%about 1000 HMXBs with luminosity larger than $10^{33}$ ergs$^{-1}$
%when we take the typical values of $\alpha=2$, $v_8=2$, and
%$\sigma=190$ kms$^{-1}$ respectively and may be detected in
%sensitive X-ray surveys.
%In Fig.~6 we show the distributions of
%their X-ray luminosities and X-ray lifetimes. Most of them can
%shine in X-rays with luminosities of $\sim 10^{32}-10^{36}$
%ergs$^{-1}$ for a few million years.
The HMXB 4U2206$+$54 is likely to be the prototype of this kind of
sources \citep{ribo05}. A number of papers have also suggestted
that some neutron stars receive low kick speeds of $\leq50$
kms$^{-1}$ at birth \citep{pfahl02, pod04, dew05}. If all the
neutron stars are born with such small kicks, our calculations
show that there would be about $4-5$ times more X-ray sources
produced.
%In this way, the assumed natal kick distribution of our
%model doesn't have significant
%effect on our results. %\citet{wan02} have reported hundreds of
%X-ray sources with luminosities $\sim 10^{33}-10^{35}$ ergs$^{-1}$
%in the the Galaxy center region, which contains $\lsim 1\%$ of the
%total Galactic population.
%\citet{pfa02} attributed a significant
%fraction of these X-ray sources to neutron stars accreting from
%the winds of the MS donor stars. Our calculations, taking into
%account neutron star spin evolution, suggest that HMXBs with MS
%donors could contribute at most a few of these sources \citep[see
%also][]{liu06}.
% However, more
%detailed investigation by \citet{liu06} shows that the number of
%these sources falls short of the observed ones if the propeller
%effect is taken in account. It remains to see how much they could
%contribute to the Galactic ridge X-ray emission along with
%coronally active binaries and cataclysmic variables \citep{saz05}.

\acknowledgments We are grateful to an anonymous referee for careful
reading the manuscript and constructive comments that significantly
improved the manuscript. This work was supported by Natural Science
Foundation of China under grants 10025314 and 10573010.

\clearpage
\begin{figure}
\plottwo{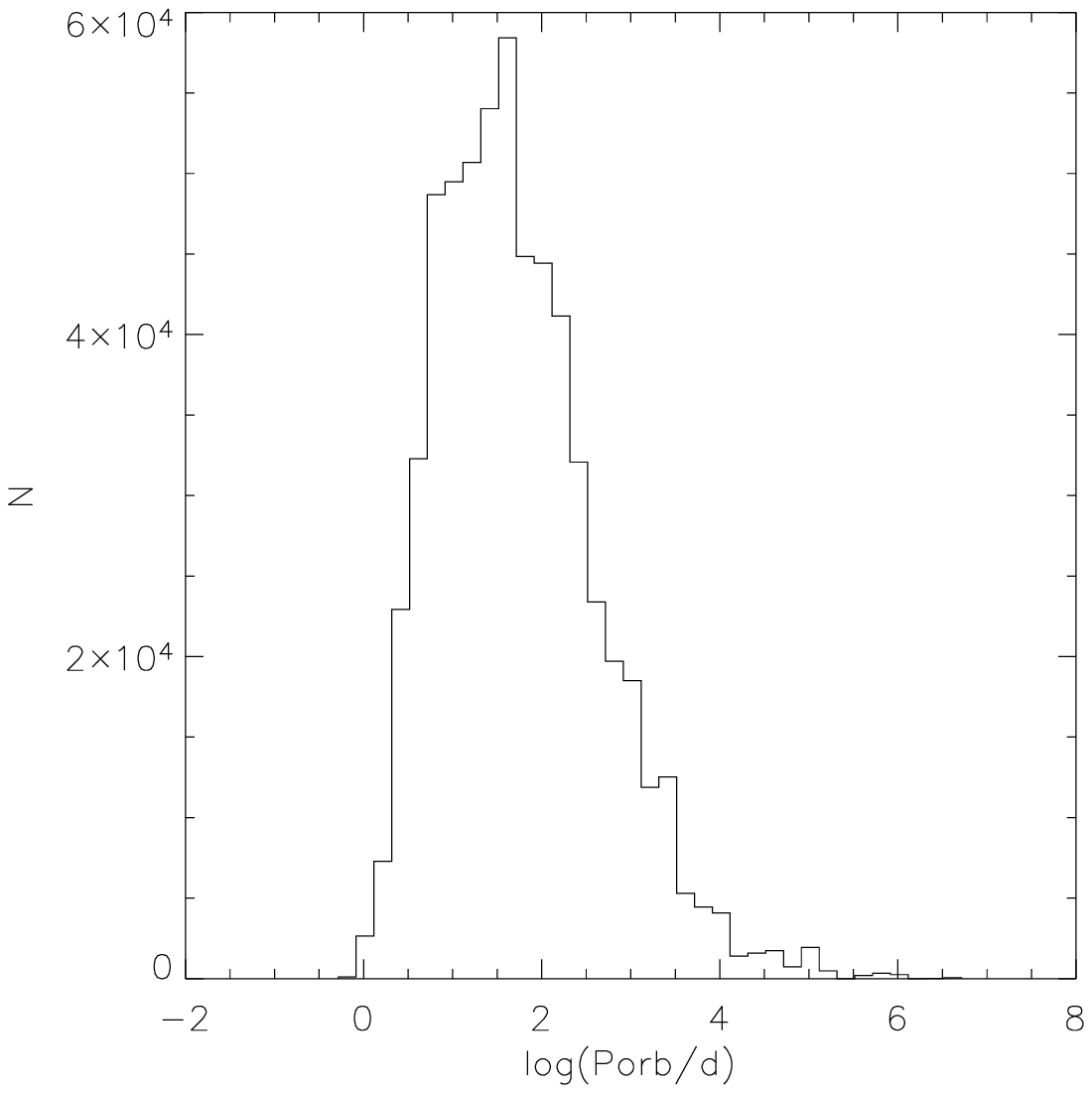}{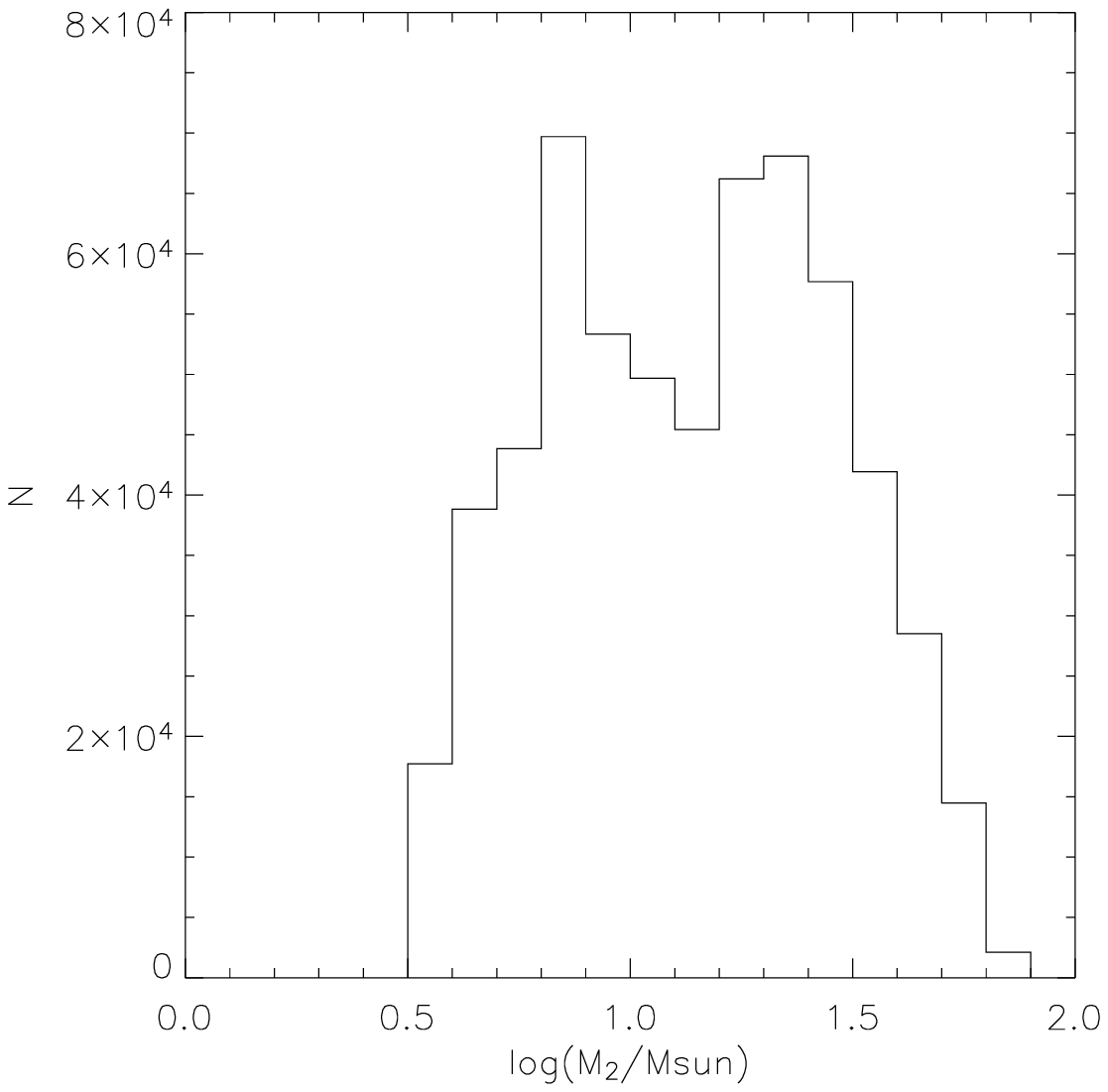}
\caption{The initial orbital period distribution of neutron star
$+$ massive MS star binaries (left) and the incipient mass
distribution of the MS companion stars (right).}
\end{figure}

%\clearpage
%\begin{figure}
%\includegraphics{fig2.eps}
%\caption{The incipient mass distribution of the MS companion
%stars.}
%\end{figure}

\clearpage
\begin{figure}
   \plottwo{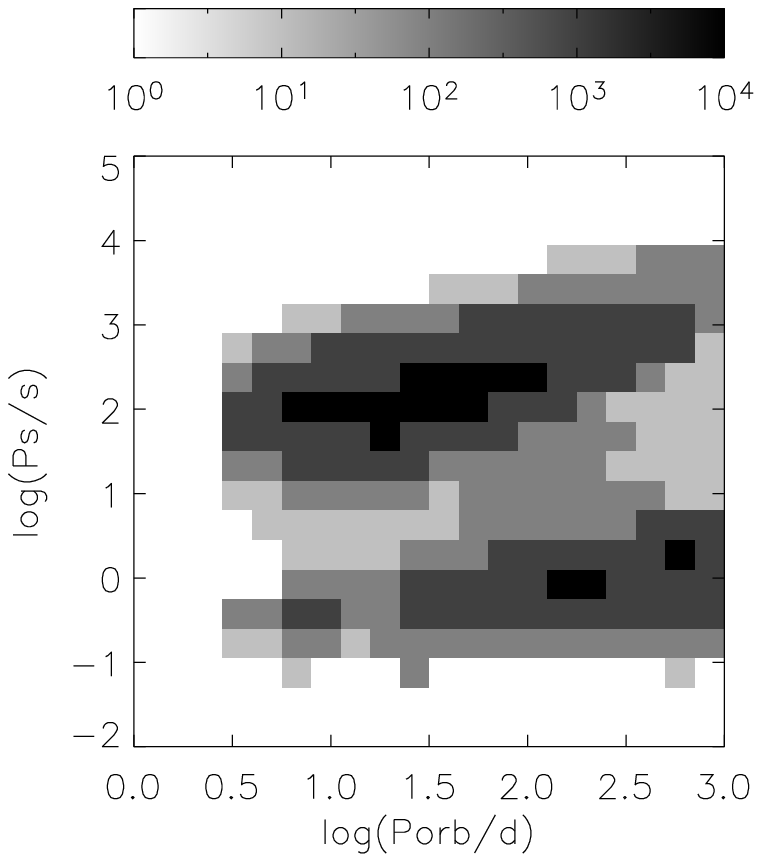}{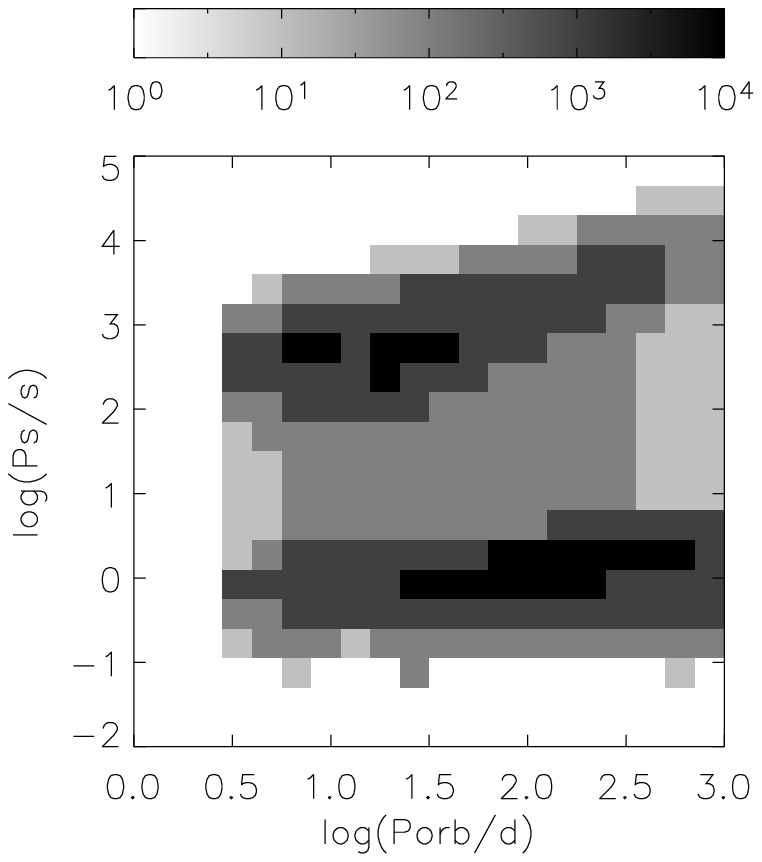}
   \caption{The distribution of neutron star binaries in the $\ps-\pb$ diagram
   when $\ps=P_{\rm eq}$ or at the age of $t_{\rm MS}$. The color bars denote
   the relative numbers in different regions. The left and right panels correspond
   to $v_8=1$ and 2, respectively.}
   \label{}
\end{figure}

\clearpage
\begin{figure}
   \plottwo{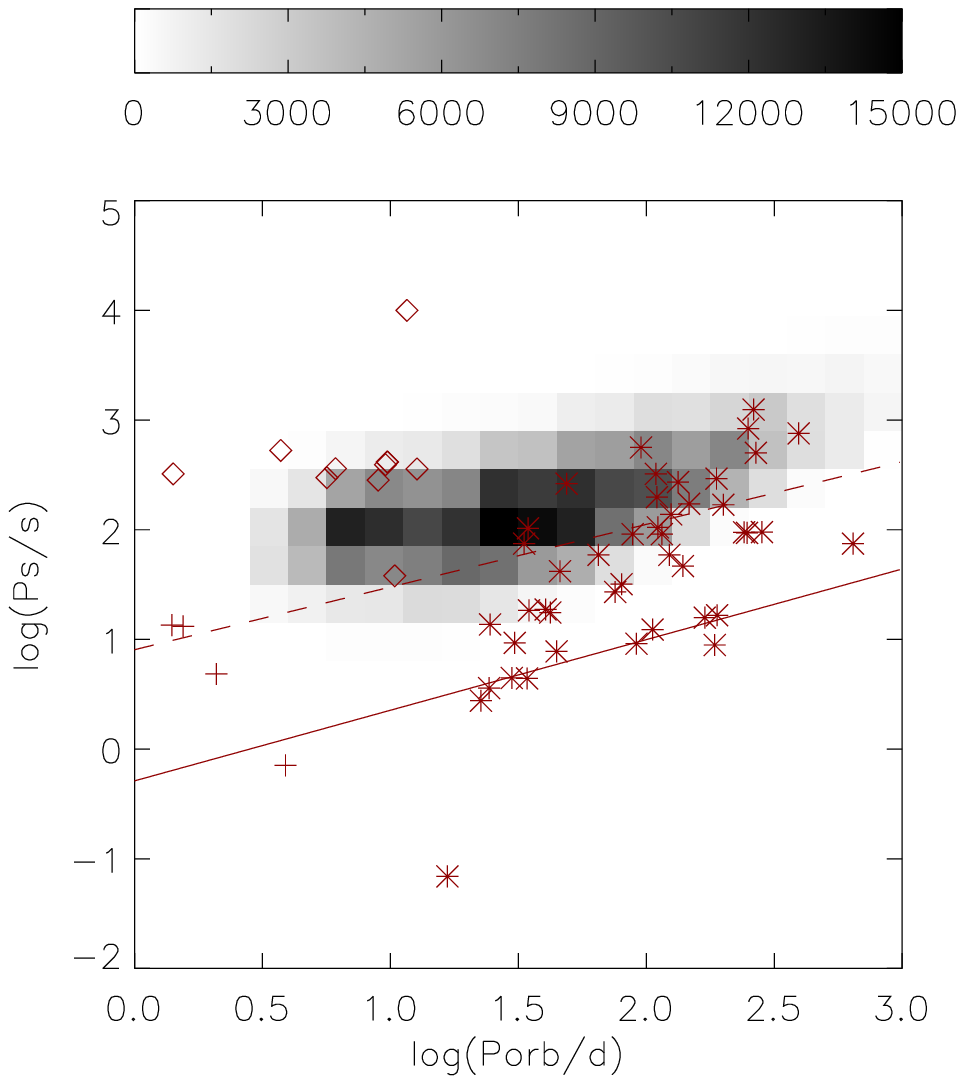}{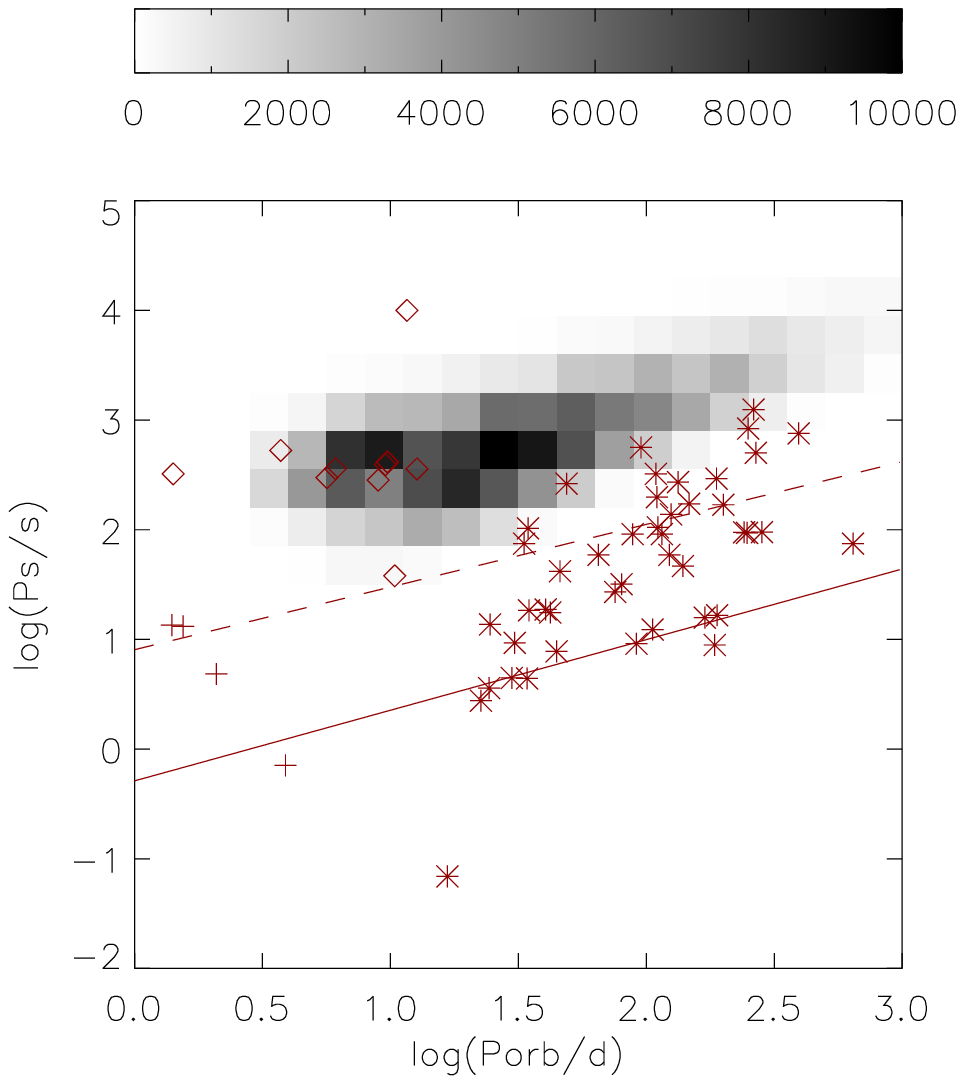}
   \caption{The $\ps-\pb$ distribution of wind-fed HMXBs. The dashed and
   solid lines represent the lower limits of the spin periods for
   supergiant and Be systems respectively. Asterisks and diamonds
   mark the Be and supergiant wind-fed HMXBs, and
   crosses for Roche-lobe overflow systems, respectively. The left and right
   panels correspond to $v_8=1$ and 2, respectively.
   }
   \label{f}
\end{figure}

\clearpage
\begin{figure}
   \plottwo{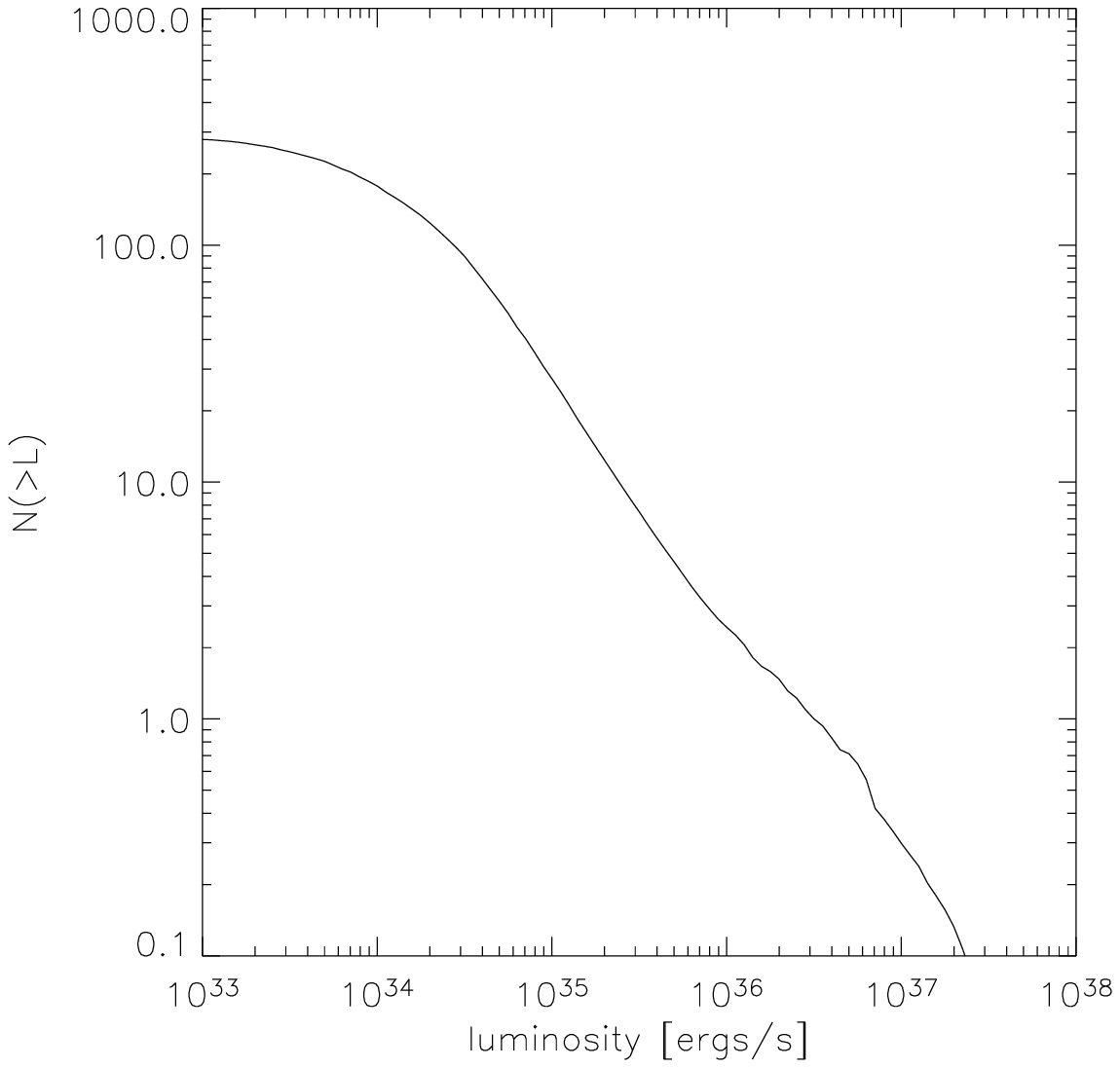}{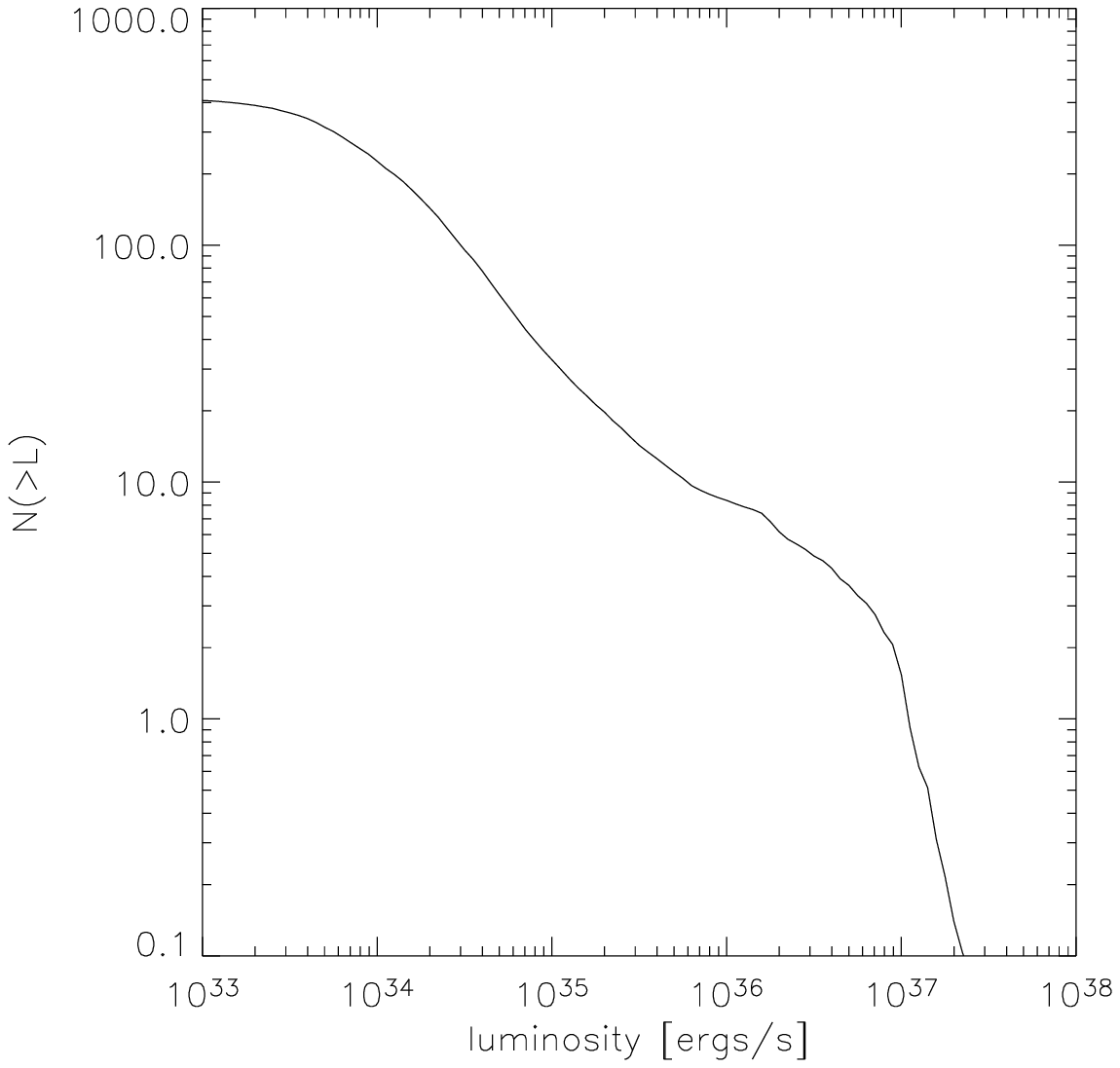}
   \clearpage
   \plottwo{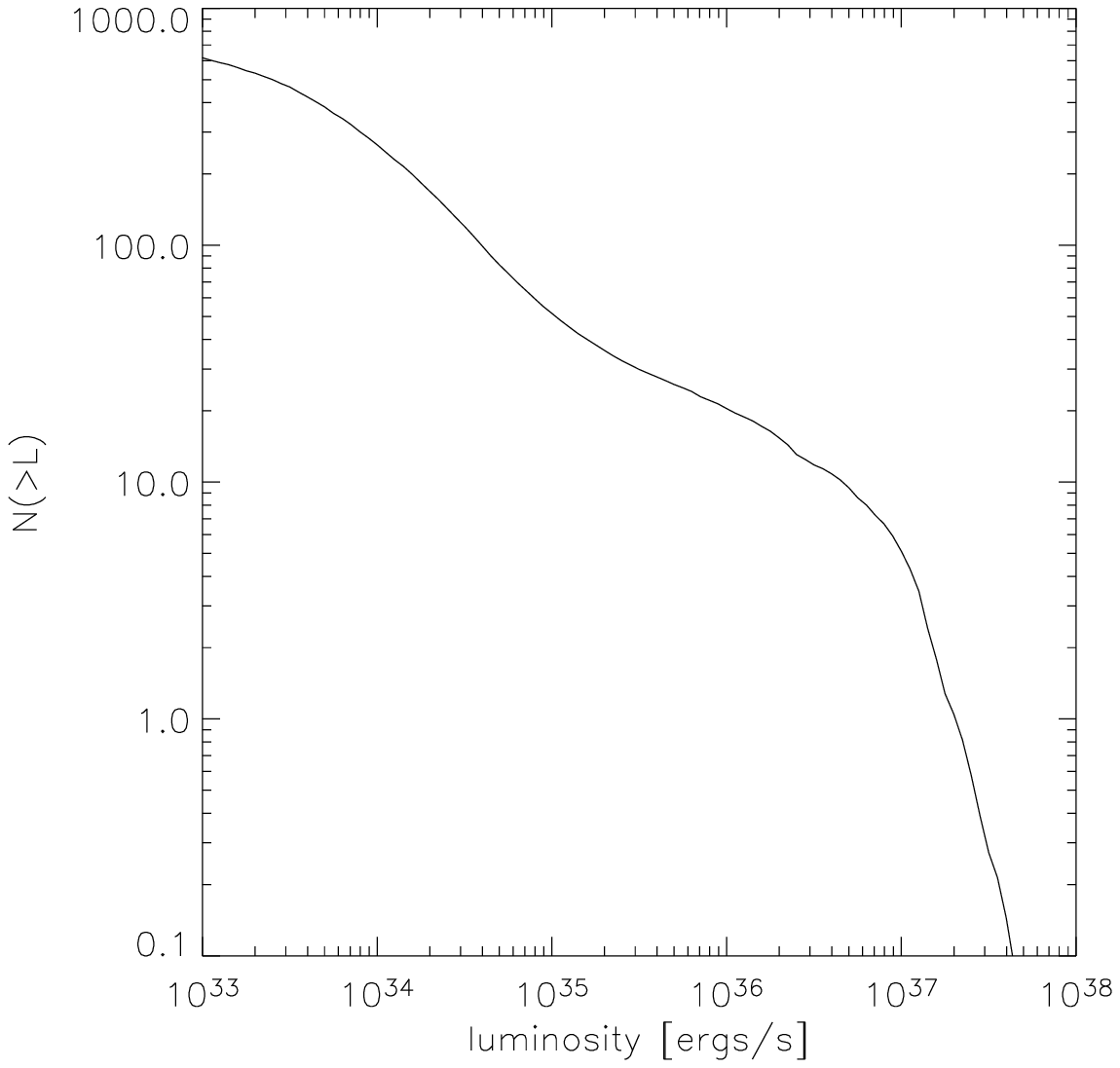}{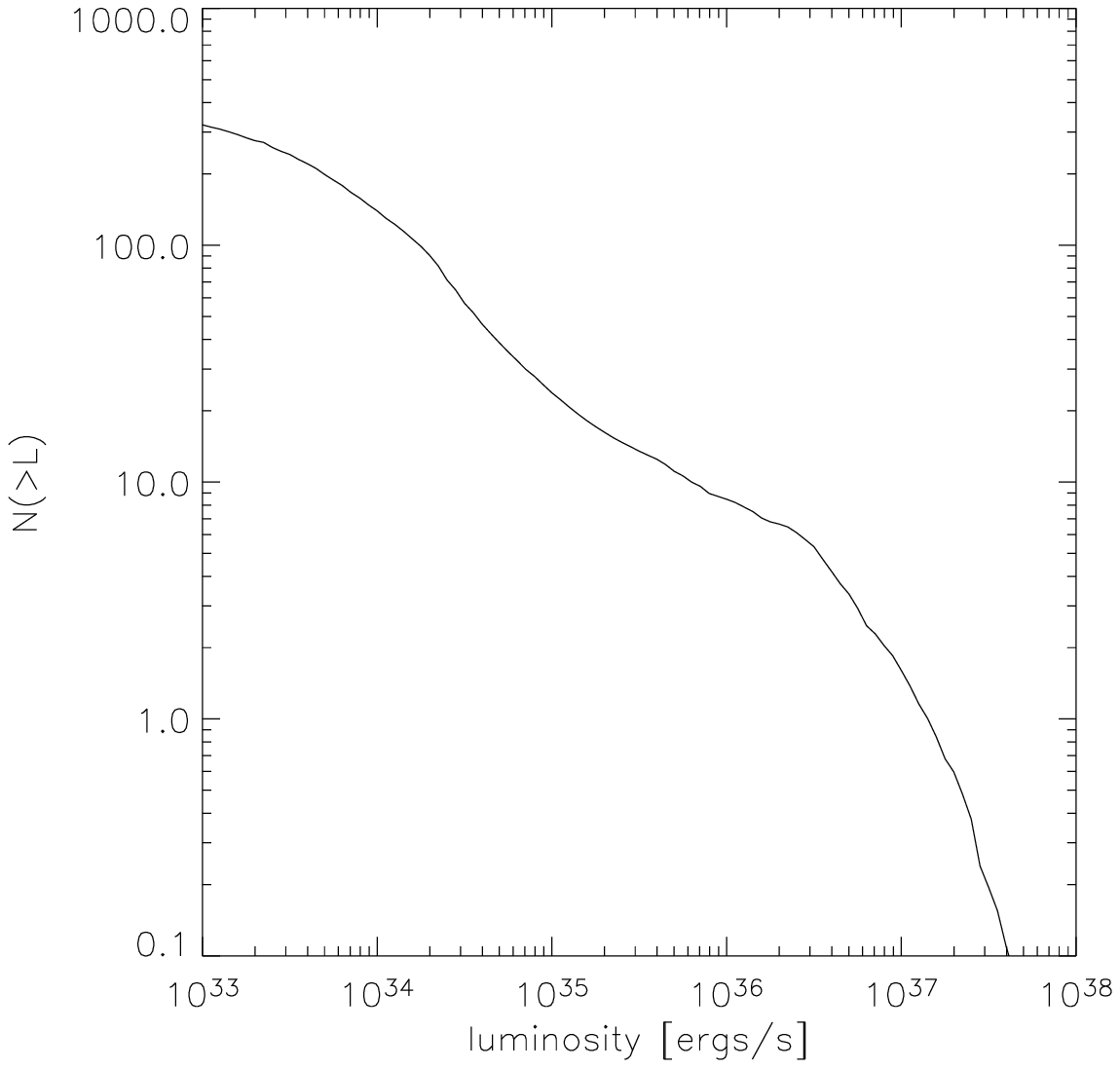}
   \caption{The luminosity distribution of wind-fed HMXBs based on
   population synthesis calculations. {\bf Top:} We take $v_8=1$, $\sigma=190$ kms$^{-1}$
   ($\alpha=0.1$ and $1$ in the left and right panels
   respectively).
    {\bf Bottom:} We take $\alpha=2$, $v_8=2$ ($\sigma=190$ kms$^{-1}$ and $265$ kms$^{-1}$  in the left
    and right panels respectively).
   }
   \label{}
\end{figure}

\clearpage
\begin{figure}
   \plottwo{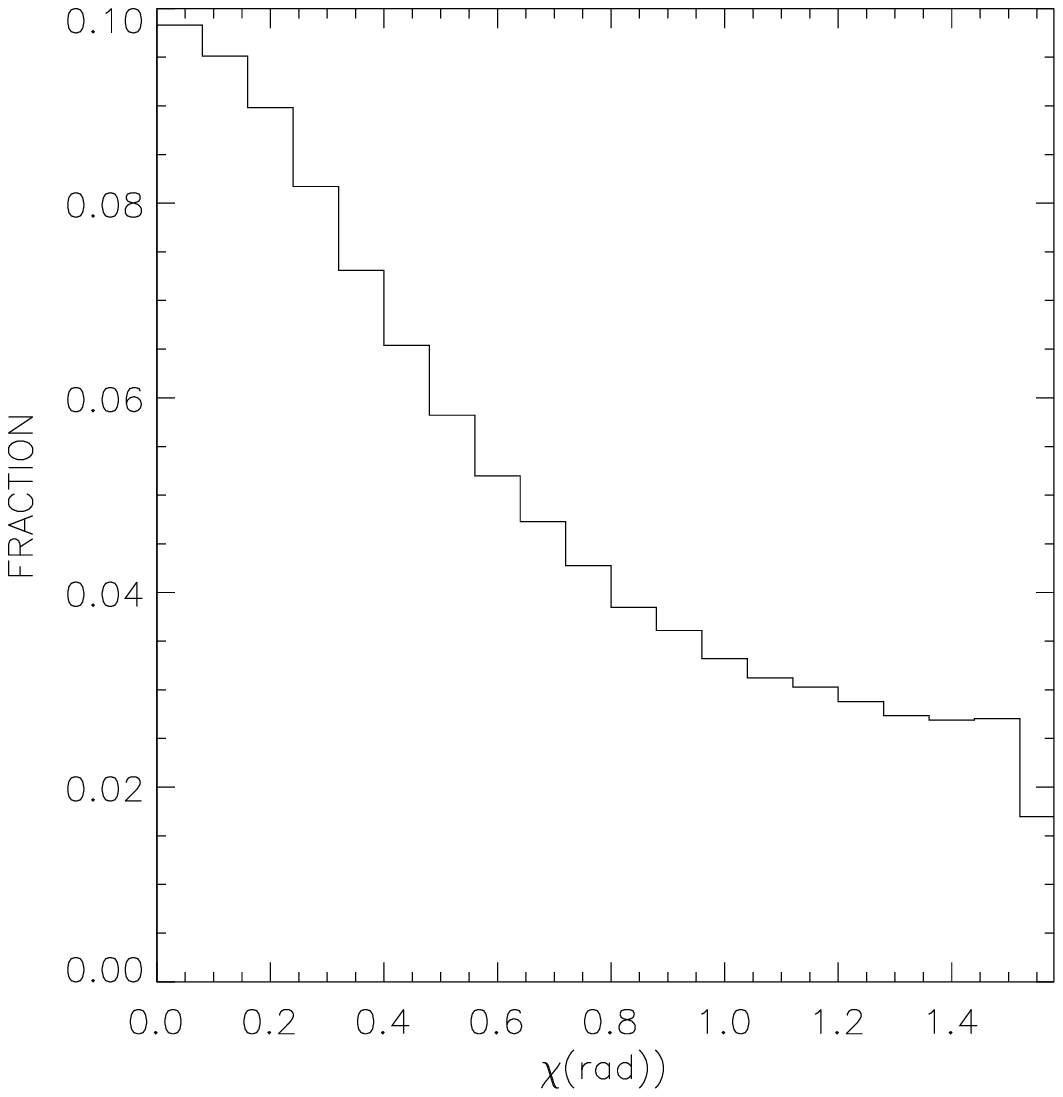}{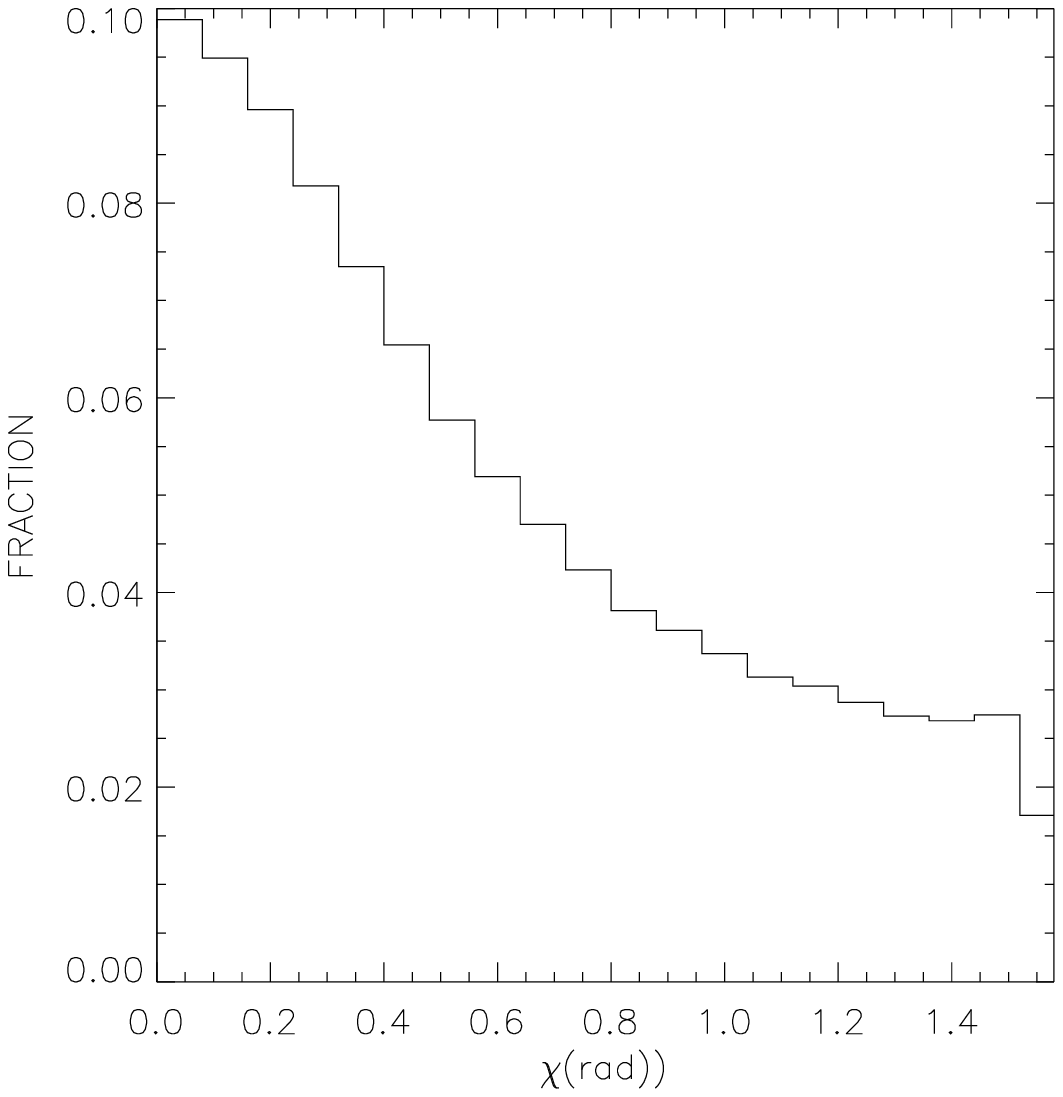}
   \caption{The relative distribution of the inclination angles of HMXBs. The left and right
   panels correspond  to $v_8=1$ and 2, respectively.
   }
   \label{}
\end{figure}

%\clearpage
%\begin{figure}
%   \plottwo{fig61.eps}{fig62.eps}
%   \caption{The luminosity and lifetime distribution of HMXBs with MS donors. The left and right
%   panels correspond  to $v_8=1$ and 2, respectively.
%   }
%   \label{}
%\end{figure}

\end{document}